\documentclass[conference]{IEEEtran}
\IEEEoverridecommandlockouts
\usepackage{cite}
\usepackage{fancyhdr}
\usepackage{amsmath,amssymb,amsfonts}
\usepackage{graphicx}
\usepackage{textcomp}
\usepackage{xcolor}
\usepackage[linesnumbered,ruled,vlined]{algorithm2e}

\SetKwInput{KwInput}{Input}
\SetKwInput{KwOutput}{Output}
\usepackage{algpseudocode}

\SetCommentSty{mycommfont}
\algnewcommand{\algorithmicforeach}{\textbf{for each}}
\algdef{SE}[FOR]{ForEach}{EndForEach}[1]
  {\algorithmicforeach\ #1\ \algorithmicdo}
  {\algorithmicend\ \algorithmicforeach}

\usepackage{url}

\def\BibTeX{{\rm B\kern-.05em{\sc i\kern-.025em b}\kern-.08em
    T\kern-.1667em\lower.7ex\hbox{E}\kern-.125em}}

\fancypagestyle{FirstPage}{
\lfoot{978-1-6654-8045-1/22/\$31.00 ©2022 IEEE} 
}

\begin{document}

\title{Learning on Health Fairness and Environmental Justice via Interactive Visualization
}
\def\plainauthor{\IEEEauthorblockN{Abdullah-Al-Raihan Nayeem$^{1}$, Ignacio Segovia-Dominguez$^{2,3}$, Huikyo Lee$^{2}$, Dongyun Han$^{4}$, \\ Yuzhou Chen$^{5}$, Zhiwei Zhen$^{3}$, Yulia Gel$^{3}$, and Isaac Cho$^{1,4}$}}
\author{\plainauthor\\ \IEEEauthorblockA{
{$^{1}$Department of Computer Science, University of North Carolina at Charlotte, Charlotte, United States} \\ 
{$^{2}$Jet Propulsion Laboratory, California Institute of Technology, Pasadena, United States} \\
{$^{3}$Department of Mathematical Sciences, The University of Texas at Dallas, Dallas, United States}\\
{$^{4}$Department of Computer Science, Utah State University, Logan, United States}\\
{$^{5}$Department of Computer and Information Sciences, Temple University, Philadelphia, United States}\\
}}

\maketitle

\begin{abstract}
This paper introduces an interactive visualization interface with a machine learning consensus analysis that enables the researchers to explore the impact of atmospheric and socioeconomic factors on COVID-19 clinical severity by employing multiple Recurrent Graph Neural Networks. We designed and implemented a visualization interface that leverages coordinated multi-views to support exploratory and predictive analysis of hospitalizations and other socio-geographic variables at multiple dimensions, simultaneously. By harnessing the strength of geometric deep learning, we build a consensus machine learning model to include knowledge from county-level records and investigate the complex interrelationships between global infectious disease, environment, and social justice. Additionally, we make use of unique NASA satellite-based observations which are not broadly used in the context of climate justice applications.
Our current interactive interface focus on three US states (California, Pennsylvania, and Texas) to demonstrate its scientific value and presented three case studies to make qualitative evaluations.


\end{abstract}


\thispagestyle{FirstPage}

\section{Introduction}

Transmissibility of various respiratory infectious diseases can be affected by weather conditions and atmospheric factors, which, in turn, have an impact on hospitalization rates. Particularly, in the context of the coronavirus disease 2019 (COVID-19) pandemic and its spread, there is no consensus on the impact of atmospheric factors on the disease dynamic and the type of relationship among 
variables~\cite{RENDANA20211320,ILOANUSI2021111340,ZHANG2022114085
}. 
Modeling and understanding the interlink between atmospheric variables and the disease spread is cumbersome while considering other factors (e.g., socioeconomic variables) also involved during infection spread. Using state-of-the-art instruments on NASA's Earth Observing System (EOS) satellites, we make the first step forward to fill in the interdisciplinary gap among exceptional NASA's satellite-based datasets, COVID-19 hospitalization dynamics, and interactive visualization systems. 

The COVID-19 pandemic provides a unique opportunity to study environmental (in)justice from both global and local perspectives, including novel dimensions of social problems that keep arising as new data and historic information comes out~\cite{EnvJusRoundtable:Sacoby:2020}. COVID-19 has unveiled the racial disparities in healthcare access and health outcomes and taught us that health fairness and environmental (in)justice goes beyond local hazards and exposure to pollutants~\cite{EnvJusClimate:Cooper:2021}. Using modern tools from artificial intelligence (AI), we can potentially find novel elements that could lead us to previously missed analysis on the role of climate in contemporary environmental (in)justice. 

In this paper, we use multiple Recurrent Graph Neural Networks (RGNNs) to forecast COVID-19 clinical severity with varying atmospheric factors. By using RGNNs, we rely on one of the most effective state-of-the-art methods to model data in non-Euclidean spaces, which are part of the broader field of Geometric Deep Learning (GDL)~\cite{bronstein2017geometric}. Given the irregular geographic structure of available data for COVID-19, and the noisy official records, GDL using the key spatiotemporal patterns in satellite observations as predictors appear to be one of the most promising forecasting approaches for tracking the hidden mechanisms behind spatiotemporal COVID-19 dynamics. Additionally, for large-scale multivariate spatiotemporal data such as our case, interactive visualizations are essential to rapidly identifying critical association patterns. The following are the key contributions of this paper:

\begin{itemize}
    \item We introduce a voting system based on multiple RGNNs to account for connectivity among US counties, which naturally introduces a transmission network of the disease at each learning stage. Our experiments take into account both the temporal and spatial dependencies simultaneously with an emphasis on forecasting the number of hospitalizations due to COVID-19.
    
    \item We provide interactive visualization that further enhances the finding of relationships between COVID-19 clinical severity and three key atmospheric factors (temperature, relative humidity, and argon oxygen decarburization (AOD)) obtained through satellite observations. 
    
    
    \item We present three usage scenarios from the interactive coordinated visualizations to discover underlying insights within the cross-domain multivariate datasets.

\end{itemize}
\section{Related Work}

\subsection{Deep Learning for Spatiotemporal Forecasting}
A natural deep learning solution to sequential data modeling consists of Recurrent Neural Networks (RNNs) and their successors~\cite{RNNreview:yu:2019}.
GNNs effectively incorporate spatial dependencies in the RNN framework, especially in traffic prediction problems on road network graphs~\cite{Yu2018ijcai}. Diffusion Convolutional RNN (DCRNN)~\cite{Li2018iclr} integrates diffusion graph convolutional layers into a GRU network for long-term traffic forecasting.
In turn, Guo et al. \cite{guo2019aaai} further employ an attention mechanism to learn spatial and temporal dependencies, while
Wu et al. \cite{Wu2019ijcai} developed a self-adaptive adjacency matrix to perform graph convolutions without a predefined graph structure.
Space-Time-Separable Graph Convolutional Network (STSGCN)~\cite{Song2020aaai} simultaneously captures the localized spatial-temporal correlations by localized spatial-temporal graphs.

\subsection{Bio-surveillance on COVID-19}

Since the outbreak of COVID-19 in December 2019, there have been countless attempts to model the spatiotemporal dynamics with particular emphasis on forecasting disease spread and mortality.  Mathematical epidemiologists study the efficacy of contemporary models~\cite{vespignani2020modelling},
and compare different intervention strategies~\cite{chang2020modelling}, whilst other researchers focus on mixed data-driven and statistical approaches such as Autoregressive integrated moving average (ARIMA) models~\cite{NESACOVID2021}, neural networks~\cite{DECARVALHO20221021}, RNNs ~\cite{ARUNKUMAR2021110861}, 
and Fuzzy Logic in combination with Particle Swarm Optimization~\cite{KUMAR2021107611}. 
However, modeling dynamics of COVID-19 clinical severity through the perspective of recorded hospitalization rates remains under-explored~\cite{segovia2021tlife,segovia2021does}. 
Our visualization framework is a novel research effort to gain an understanding of the non-linear dynamics of COVID-19 clinical severity in the presence of atmospheric and socioeconomic factors.

\subsection{Visualization Methods}


Numerous visualization systems are developed for COVID-19 monitoring \cite{johnshopkins, who, uw} that provide the confirmed cases, active cases, recoveries, and deaths. For instance, CoronaViz \cite{samet2020} is a visualization system with an interactive view and a data control panel to unfold patterns in the spatiotemporal COVID-19 variables. The data control in CoronaViz provides location and feature-based filtering as well as an option to normalize based on population. Geo-OEDV \cite{bernasconi2021} 
is a visual framework for the exploratory analysis of clinical, socioeconomic, and environmental data for COVID-19 targeting the general users, policymakers, and researchers. Another visual analytics environment is proposed for public health officials to perform exploration and decision-making tasks with COVID-19 clinical data \cite{afzal2020}.
For multivariate data analysis, CrossVis \cite{Steed2020} demonstrated an interactive parallel coordinates plot consisting of categorical and numerical axes to explore large-scale heterogeneous spatiotemporal data. Multifaceted parallel coordinate plots are utilized in other research to illustrate the association of spatiotemporal variables~\cite{wang2018b, li2020sovas}.

To our best knowledge, none of the existing visualization systems supports incorporating the cross-domain data for analyzing the impact of COVID-19 clinical severity factors. Hence, our visualization that accommodates multivariate cross-domain data can benefit researchers to include variables from fragmented data sets in forecasting the clinical severity.

\section{Datasets}\label{sec:dataset}
Our research utilizes three real-world heterogeneous datasets: the COVID-19 dataset, the NASA dataset, and the socioeconomic dataset. 

\textbf{COVID-19 dataset. } Since our approach produces COVID-19 hospitalization forecasts at the county-level resolution, in accordance with data availability, we focus our experiments on three US states: California (CA), Pennsylvania (PA), and Texas (TX). Daily records on COVID-19 and hospitalizations are taken from curated datasets at the CovidActNow project\cite{uscovidtracker}. 
We follow the disease progression and guidelines from the Models of Infectious Disease Agent Study (MIDAS) for COVID-19 modeling research~\cite{midas}.
We use time series, with values obtained on a daily basis, from February 1 to December 31, 2020, thus focusing on the period of time when hospitalization numbers were publicly available for all three states, at a county level.

\textbf{NASA dataset. } To avoid current limitations on ground-based climate observations, in terms of resolution on covered areas across the US, we rely on processed satellite-based observations of temperature, relative humidity, and AOD. NASA’s Distributed Active Archive Centers (DAAC) servers provide publicly available access to the original datasets through the products: 1) Atmospheric Infrared Sounder (AIR)/Aqua L3 Daily Standard Physical Retrieval (AIRS3STD) \cite{airs}
for temperature and relative humidity, and 2) the Atmosphere Daily Global Product from Moderate-resolution Imaging Spectroradiometers (MODIS) on Terra (MOD08\_D3) \cite{modis}
for AOD. Our climatology datasets (i.e. temperature and relative humidity variables) come from averaging 6209 days from January 1, 2003 to December 31, 2019. Thus, we compress annual cycles to build the whole annual cycle of 2020, using the level 3 product (L3) data from regular Gaussian grids with $1^{\circ} \times 1^{\circ}$ resolution. Similarly, our AOD dataset is calculated using AOD at 550 $nm$ wavelength from the 6939 days from January 1, 2001 to December 31, 2019; thus producing an average annual cycle. 
We match each county-level time series with its corresponding Federal Information Processing Standard Publication 6-4 (FIPS 6-4) code. Thus, our datasets are easier to use along with other datasets in the same county-level resolution; e.g., socioeconomic indicators, infectious diseases, transportation, etc.


\textbf{County-level connectivity. } We analyze the impact of two types of graph connectivity: 1) border-based county connections, and 2) socioeconomic-based county connections. Network connectivity, using shared borders, is built via the official county adjacency file record layout
from the US Census Bureau. 
We generate a socioeconomic-based network for each state under study using 9 socioeconomic variables. 
A socioeconomic matrix distance, via the euclidean metric, helps us select the strongest county-level connections which serve as input to each GNN model. Five variables come from the Centers for Disease Control and Prevention  (CDC)/Agency for Toxic Substances and Disease Registry (ATSDR) Social Vulnerability Index \cite{cdc}: Socioeconomic Status, Household Composition \& Disability, Minority Status \& Language, Housing Type \& Transportation, and Overall Vulnerability Index. The rest of the variables are part of The COVID-19 Vaccine Coverage Index (CVAC) \cite{cvac}: Historic Undervaccination, Sociodemographic Barriers, Resource-Constrained Healthcare System, and Healthcare Accessibility Barriers. 


\section{Design Method}\label{Sec:DesignMethod}
To evaluate the overall performance of DL/GDL models on spatio-temporal forecasting tasks, in our experiment, we use 8 DL and GDL models to forecast the next $\omega$ steps data based on past $h$ steps historical data, i.e., $\{X_{t+1}, X_{t+2}, \dots, X_{t+\omega}\} = \mathcal{M}(X_{t-h+1}, X_{t-h+2}, \dots, X_t)$, where $\mathcal{M}(\cdot)$ denotes DL/GDL model and $X_{i}$ is the node features of graph $G_t$ at timestamp $i$ ($i \in \{0, 1, \dots, \tau\}$).
\begin{figure}[t]
    \centering
    \includegraphics[width=\linewidth]{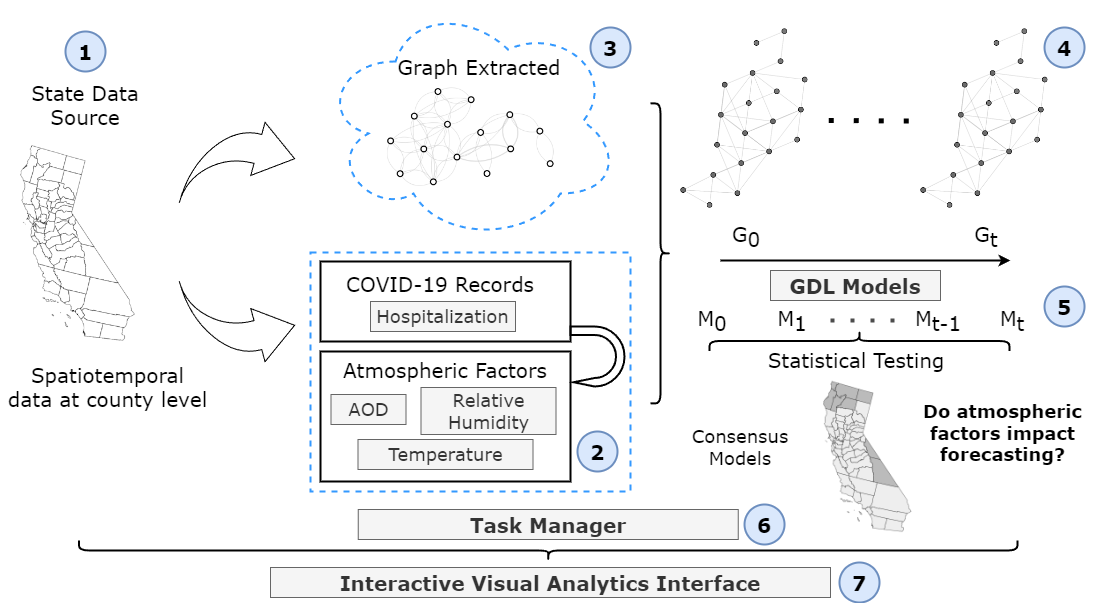}
    \caption{The system pipeline for GDL models and interactive visualization interface for county-level geographical model consensus of COVID-19 clinical severity risk.}
    \label{fig:system-pipeline}
\end{figure}  


\textbf{Benchmark models}. We benchmark two broad classes of neural networks 
(i) RNNs: LSTM can forecast multivariate time series with LSTM hidden units; (ii) Spatiotemporal Graph Convolutional Networks: spatiotemporal model with the framework of graph convolutional network (GCN) exploit GCN and temporal convolution to capture dynamic spatial and temporal patterns and correlations; we report performances for 8 types of state-of-the-art methods on our benchmark datasets: 
LSTM, DCRNN, LSTM R-GCN, Attention Temporal GCN (A3T-GCN), Message Passing Neural Networks with LSTM (MPNN+LSTM), Evolving GCNs (EvolveGCNO), and Gated Graph Neural Network (GG-NNs) for Dynamic Graphs (DyGrEncoder).

\textbf{Experimental setting for forecasting.} We use daily data from February 1 to December 31, 2020, and split the graph signals into a train set (the first 80\% of days) and a test set (the last 20\% of days). The training step uses 5 lags of daily reported values to produce 15 days ahead forecasting. We train each model using a hidden layer dimension of 128, and output dimension of $\{55, 60, 251\}$ for CA, PA, and TX, respectively; as well as 500 epochs, a dropout rate of 0.5, and an AMSGrad optimizer using 0.02 as a learning rate. We execute 10 runs for each GDL model on each state to report prediction results with statistical significance at both state and county levels.

\textbf{GDL models for spatiotemporal forecasting}. We present the recurrent graph convolution operation on spatiotemporal networks in Algorithm~\ref{GDL_algorithm}. Given the time-varying networks (i.e., ${\bf{\boldsymbol{X}}} = \{X_0, \dots, X_t, \dots, X_\tau\}$ and ${\bf{\boldsymbol{G}}} = \{G_0, \dots, G_t, \dots, G_\tau\}$), for each timestamp $t$, we first operate graph convolution on each node $v$ and obtain graph embedding $H_t$, then feed the output embedding into RNN layer (e.g., Gated Recurrent Units (GRU)) for temporal dynamics modeling. In Algorithm~\ref{GDL_algorithm}, $f_\textnormal{MLP}$, $\textnormal{AGG}(\cdot)$, $\textnormal{CONCAT}(\cdot)$, $\phi(\cdot)$ denote the multilayer perceptron, aggregation function (e.g., average, sum, max), concatenation operation, and nonlinear activation function (e.g., {\it ReLU}) respectively; $W_g$, $W_z$, $W_r$, $W_q$, $b_z$, $b_r$, $b_q$ are trainable weights, and $\odot$ denotes the element-wise product.

\begin{algorithm}[t!]
\SetAlgoLined
\KwInput{${\bf{\boldsymbol{X}}} = \{X_0, \dots, X_T\}$ and ${\bf{\boldsymbol{G}}} = \{G_0, \dots, G_\tau\}$}
\KwOutput{${\bf{\boldsymbol{\Tilde{Q}}}}_t = \{\Tilde{Q}_1, \Tilde{Q}_2, \dots, \Tilde{Q}_t, \dots, \Tilde{Q}_\tau\}$}
\tcc{Graph convolution operation on $\{G_t, X_t\}$}
\For{\textnormal{the graph} $\{G_t, X_t\}$ \textnormal{at timestamp $t$ in mini-batch}}{
$h_t^{(0)} = f\textsubscript{MLP} (X_t)$\\
\For{$\ell \gets1$ \KwTo {\it nlayers}}{
$h^{(\ell-1)}_{t, \mathcal{N}(v)} = \textnormal{\small AGG}\left(h^{(\ell-1)}_{t, u}, \forall u \in \mathcal{N}(v) \right)$\\
$h^{(\ell)}_{t, v} = \sigma\left(W_g \cdot \textnormal{\small CONCAT}\left(h^{(\ell-1)}_{t, \mathcal{N}(v)}, h^{(\ell-1)}_{t, v}\right)\right)$\\
}
}
$H_t = \textnormal{Readout}(\{h^{(\ell)}_{t,v}, \forall v \in \mathcal{V}\})$ 
\tcc{Embedding} 
\tcc{Recurrent neural network layer}
$Z_t = \phi(W_z[H_{t-1}, H_t] + b_z)$\\
$R_t = \phi(W_r[H_{t-1}, H_t] + b_r)$\\
$\hat{Q}_t = \tanh{(W_q[R_t \odot \hat{Q}_{t-1}, H_t]) + b_q}$\\
$\Tilde{Q}_t = Z_t \odot \hat{Q}_{t-1} + (1-Z_t) \odot \hat{Q}_{t}$\\
\caption{Recurrent graph convolution operation~\label{GDL_algorithm}}
\end{algorithm}

\textbf{Consensus analysis}. We compute forecasting using 8 state-of-the-art models. 
Our consensus analysis consists in adding votes, at a county level, whenever an RNN/GNN model improves its results (when adding an atmospheric variable) with respect to the baseline (removing the atmospheric variable). Each improvement is validated through one-tailed hypothesis testing, at \textit{significant} level, on reduction of the root-mean-square error (RMSE). To the best of our knowledge, there are no previous attempts to use consensus among multiple RNN/GNN models to characterize the impact of atmospheric factors on COVID-19 clinical severity, along with an interactive visualization system to complement the decision-making of experts. 






\textbf{Analysis Pipeline.}
Fig.~\ref{fig:system-pipeline} illustrates our data analysis workflow. First, we collect and pre-process the spatiotemporal data, both COVID-19 hospitalization numbers, and satellite-based observations of temperature, relative humidity, and AOD. Next, we generate two types of county-level connectivity: 1) geographic-based border county connections, and 2) socioeconomic-based county connections. Using a chosen graph connectivity and dynamic measurements at each node, we build a dynamic node-feature network $\{G_0, ..., G_{\tau-1}, G_{\tau}\}$ for the GNNs models. Our platform allows us to analyze results using both types of networks by using these as input in our GDL models. We train a set of models $\{M_0, ..., M_{\tau-1}, M_{\tau}\}$, as described in Section~\ref{Sec:DesignMethod}, to forecasting hospitalization numbers with and without atmospheric factors. Next, we statistically verify, and count, any model that produces an improvement in forecasting results when adding atmospheric factors. This consensus model produces geographical patterns which help us identify communities with a higher risk of hospitalization due to COVID-19. 


Note that this same framework can be used to study other infectious diseases, such as Zika or Monkeypox, which in turn will impact government policies and health decision-making.

\begin{figure*}[t]
    \centering
    \includegraphics[width=\linewidth]{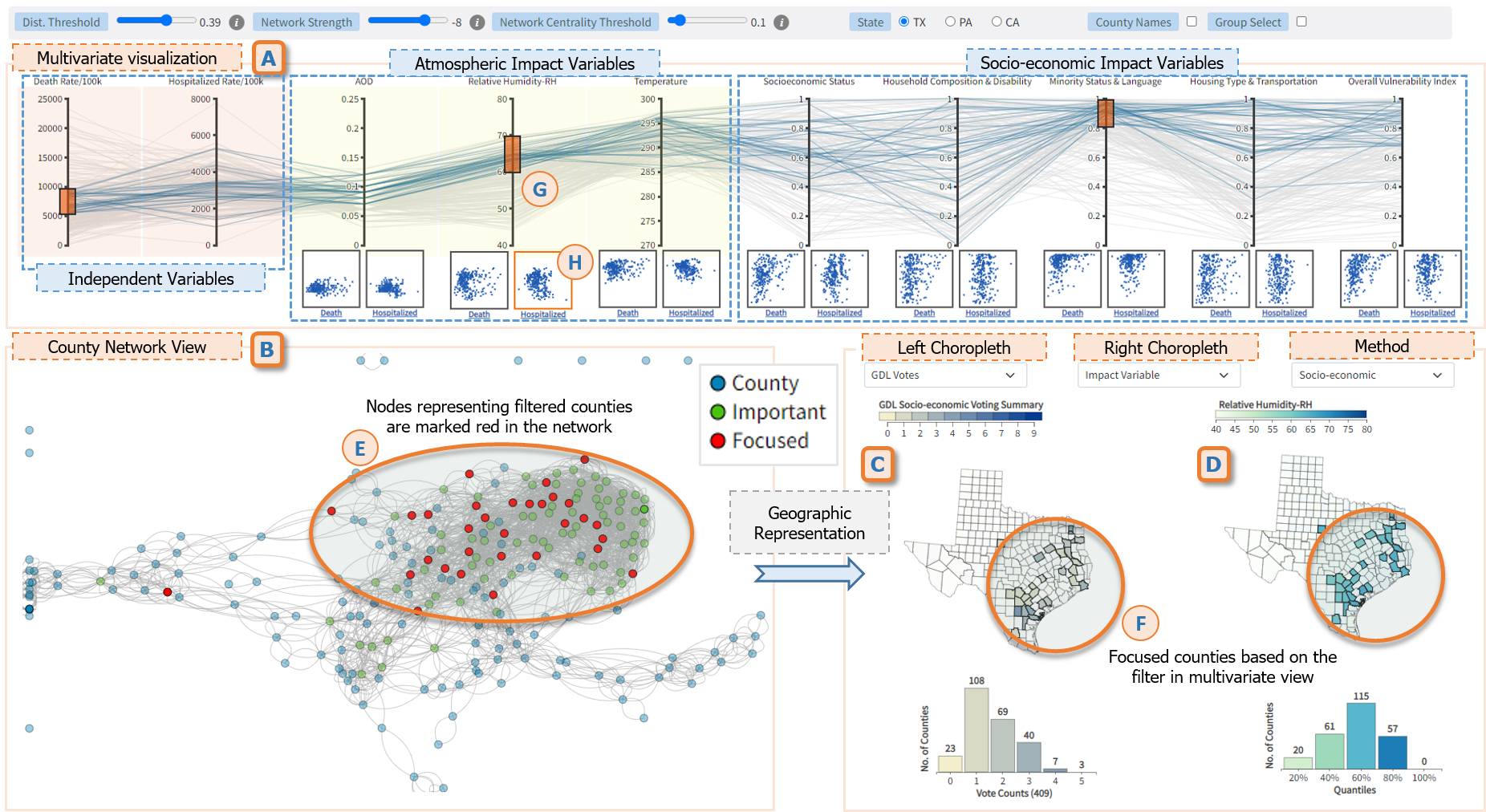}
    \caption{Overview of the interactive visualization interface. A) Parallel coordinates presenting a multivariate visualization of clinical, atmospheric, and socioeconomic variables of the counties. B) A county network illustrates the relationships among the counties based on socioeconomic similarities. C) Choropleth map views presenting the geographical representation of the counties with the summary of participating variables and the GDL voting result. A control bar at the top supports reconfiguring these interactive visualizations.}
    \label{fig:teaser}
\end{figure*}

\section{Interactive Visualizations}
The visual interface presents interactive and coordinated visualizations to facilitate the user for exploring the clinical, atmospheric, and socioeconomic variables as well as the voting results from GDL models. To support the interactive analysis of both input and resulting variables, the interface consists of four main views: A) Multivariate Visualization to explore the input data variables (e.g., death rate, hospitalization rate, socioeconomic status), B) County Network View to explore the connectivity among the counties of a specific US state, C) Choropleth Map View to provide state choropleth maps to explore the summarized data variables for each county by its geographic location, and D) Voting Results View to explore the GDL results to determine the impact of socioeconomic variables on COVID-19 clinical severity factors. The interactive visualizations are served via a web-interface.\footnote{The visual interface is available online: https://vizus.cs.usu.edu/app/gdlviz/} 



\textbf{Multivariate Visualization.}
To examine the impact factors, we categorize the input variables between two groups: primary and impact variables. The primary variables are death and hospitalization rates. Since the county population has a substantial difference among others, we normalized the primary data variables to find the rate per 100K population. Moreover, the impact variables have the five variables from the CDC/ATSDR social vulnerability index. To support the investigation of correlation among the primary and impact data variables, we leveraged parallel coordinates that are widely used visualization for multivariate data \cite{Steed2020}. 

The dimensions in the parallel coordinates combine the primary and impact variables. Each axis in the parallel coordinates represents a dimension. Since we are investigating the impact variables against the primary variables, the axes for the primary variables are highlighted in orange color (Fig. \ref{fig:teaser}A). Each connecting line in the parallel coordinates represents a county of the selected state. Hence, the total number of connecting lines in the parallel coordinates matches the number of counties in the state. Each county line connects the dimensions based on its corresponding value. The parallel coordinates provide interactive data filtering to specify the window size for each axis (Fig. \ref{fig:teaser}G). The filtered county lines are highlighted in blue color while other lines are in gray color. 

An axis for each impact variable includes two mini scatter plots on the bottom (Fig. \ref{fig:teaser}H). The left and right scatter plots depict a glimpse of the corresponding correlation with the death and hospitalization rates respectively. The interface provides a detailed view of each scatter plot including a trend line for the selected variables on clicking a mini scatter plot. 

\begin{figure}[t]
    \centering
    \includegraphics[width=\linewidth]{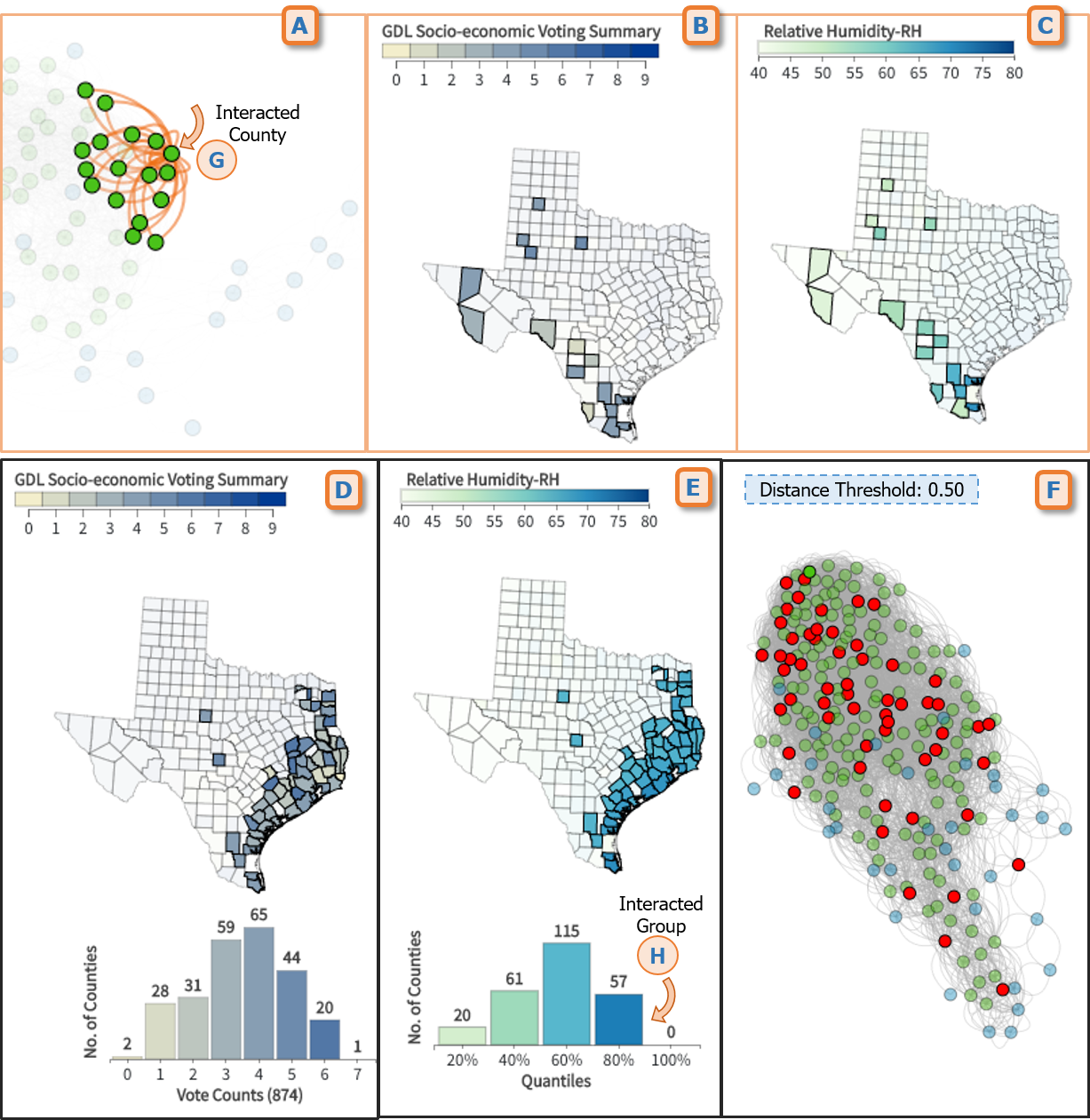}
    \caption{The interactive interface demonstrates exploratory user interactions that coordinate multiple views facilitating visual sense-making.}
    \label{fig:user-interaction}
\end{figure}

\textbf{County Network View.}
This view presents each county as a node 
and the connections between the two counties are represented with an edge. The length of the edge denotes the euclidean distance (i.e., socioeconomic similarity) between the two connected counties. County nodes within the network can be focused or selected with mouse interaction. Hovering a mouse cursor on a county node shows all neighboring counties in the network as shown in Fig. \ref{fig:user-interaction}A. Moreover, we calculate network centrality to find the significant counties in the network. Network centrality provides a tool to rate/rank the nodes in the network based on the number of degrees (edges) \cite{ibarra1993network}.

The view has a control bar to change the distance threshold, network centrality threshold, network strength, etc. The distance threshold filters out the connections that are far from the selected value. Therefore, the higher value of the threshold provides a dense-edge network as shown in Fig. \ref{fig:user-interaction}F. The network centrality threshold provides control to the user to determine the rank where a county is to be considered statistically significant. Initially, all the counties are marked blue whereas the important counties based on the network centrality threshold are marked green (Fig. \ref{fig:use-case-1}). 

\textbf{Choropleth Map View.}
This view is designed to present a state's county-level information on independent and impact variables as well as to depict the patterns based on geographical features. We utilized different color scales associated with the data variables to illustrate the value for each county. The view has two choropleth maps (Fig. \ref{fig:teaser}C and D) to provide a comparative capability to the user. These reconfigurable choropleth maps are labeled as left and right choropleth where the user can select their variable of interest. 

The data presented in the choropleth map are categorical, regardless of independent, impact, or resulting variables. For the continuous variables such as hospitalization rate per 100k and relative humidity, we calculated the quantiles to categorize the counties. For example, Fig. \ref{fig:user-interaction}F suggests that counties that lie in the top 80\% intensity of relative humidity are located on the east side of TX. Choropleth maps are also employed to visualize the summary of the voting results discussed in Section \ref{Sec:DesignMethod}. Since we used 8 GDL models, each county can get 0 to 8 votes from the model output. When the user hovers over a county, this view provides additional information on a tooltip such as the county name with code, values for the corresponding atmospheric and socioeconomic variables, and a breakdown of the forecasts predicted by the individual deep-learning models.

A histogram is implemented below the choropleth map to summarize the categorical data presented on the map. The bars in the histogram used the same color scale used for the choropleth map. The histogram presents the number of counties for each category (e.g. vote count, and quantile) as shown in Fig. \ref{fig:user-interaction}D-F. Hovering a mouse cursor over a bar highlights the corresponding counties on the choropleth map. 

\begin{figure*}[t]
    \centering
    \includegraphics[width=\linewidth]{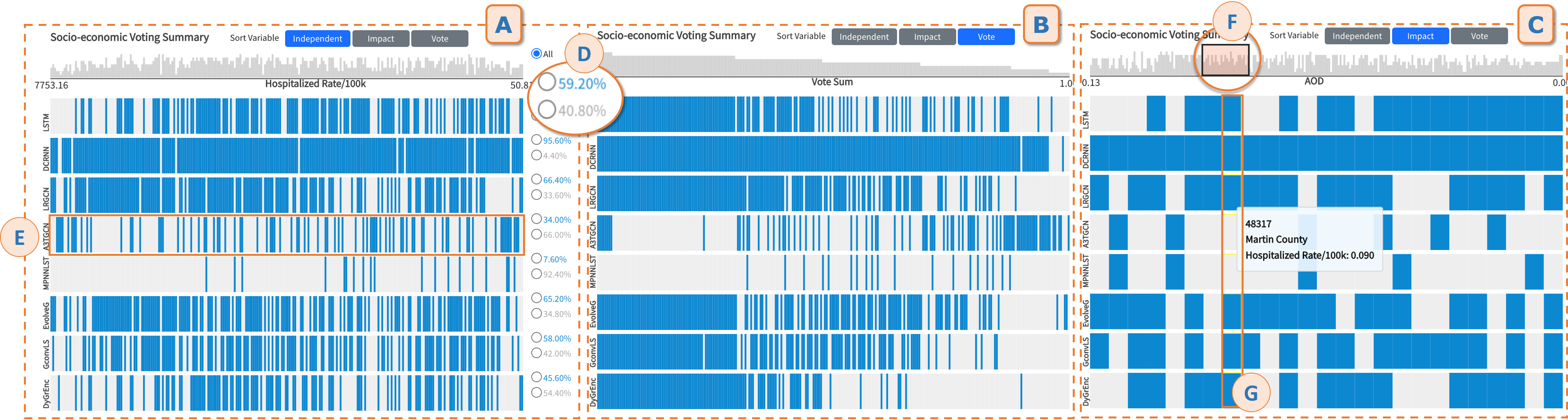}
    \caption{Pixel bar view provides a detailed illustration of the voting results from GDL models. Interactive sorting and filtering strategies are implemented to support the exploration and logical reasoning of counties' vote counts for individual states.}
    \label{fig:voting-result}
\end{figure*}
\textbf{Pixel Bar View.}
A compound view that consists of two visualizations combined with interactive features - a) a bar chart containing the vote count for the counties (Fig. \ref{fig:voting-result}A) where each bar represents a county in the selected state, b) followed by the pixel bar view that provides a breakdown of the voting results (Fig. \ref{fig:voting-result}E). Both visualizations share the same x-axis and are vertically aligned, combining 8 rows, each representing a GDL model. The pixel bar view marks the county's associated pixel with blue color if the corresponding model results in a positive prediction.

This view provides interactive features to allow the user to reconfigure the rendered visualizations based on data sorting and filtering. We implemented a couple of sorting strategies to support the user in identifying patterns. The sorting strategies are applied by the participating independent variable (e.g., hospitalization rate/100k), impact variable (e.g., AOD), and vote count for a county (Fig. \ref{fig:voting-result}). The sorting determines the order of the counties presented in the pixel bar view. For example, Fig. \ref{fig:voting-result}B illustrates the pixel bar sorted by the vote count where the counties with similar vote counts are tied together. This feature provides an idea of how individual models performed and whether there is any bias in the vote count. The user can also filter the view while interacting with the bar chart at the top (Fig. \ref{fig:voting-result}F). When the user selects a group of neighboring counties ordered in the bar chart, the pixel bar view is reconfigured to present the selected counties. Therefore, the user can focus on a specific group and identify the county names interacting with the pixel bar (Fig. \ref{fig:voting-result}G). The selection works as a sliding window where the view allows the user to shift left/right and change the window size. 

\textbf{User Interactions.}\label{sec:user-interactions}
We implemented user interactions based on data and view manipulation capabilities to facilitate the human cognition process. The user interactions are categorized based on their support for different visual analysis tasks \cite{lu2017state}. In our visual interface, we leveraged select, explore, filter, reconfigure, and connect interactions.

The user can switch a state (TX, PA, or CA) from the control bar. 
In addition, the user can specify by interacting with the scatterplots on the multivariate parallel coordinates. For example, selecting a scatterplot presenting the hospitalization rate and AOD (Fig. \ref{fig:teaser}H) reconfigures the context for the choropleth views. 
Selection is also utilized for specifying the independent and impact variable while interacting with the scatterplots on the multivariate parallel coordinates. For example, selecting a scatterplot presenting the hospitalization rate and AOD (Fig. \ref{fig:teaser}H) reconfigures the context for the choropleth views. 
The user can switch between pre-computed model results based on the socioeconomic distance and boundary-based output of the GDL models. Moreover, hovering a mouse over a county on the vote-count choropleth map view shows the GDL models that contribute to that count. Interacting with a node in the county network also highlights the neighboring counties (Fig. \ref{fig:user-interaction}) based on the selected distance threshold. 
Filtering interaction allows the user to focus on a specific subset or area of interest. The parallel coordinates presenting multivariate data support filter interaction for each data axis as shown in Fig. \ref{fig:teaser}A. The county network view also provides a group select option where the neighboring counties can be identified as a cluster on the network (Fig. \ref{fig:teaser}E). Previously, we also discussed the filtering features for the pixel bar view and choropleth map view interacting with the histogram.

The visual interface demonstrates a connected and coordinated response to the user's filter interactions. For example, in Fig. \ref{fig:teaser}A, the filter applied on the multivariate parallel coordinates essentially triggered a filtering effect on the other visual components such as Fig. \ref{fig:teaser}B-C-D. Similarly, interacting with a node on the network or a bar chart imposes a coordinated effect on all other views as shown in Fig. \ref{fig:user-interaction}.





\section{Usage Scenarios}
We present three usage scenarios of our proposed visual interface. For each usage scenario, we focus on presenting the coordinated visualization's capabilities to support exploratory analysis and provide logical reasoning to the discovered knowledge to demonstrate the scalability and adaptability of our interface.

\begin{figure}[t]
    \centering
    \includegraphics[width=\linewidth]{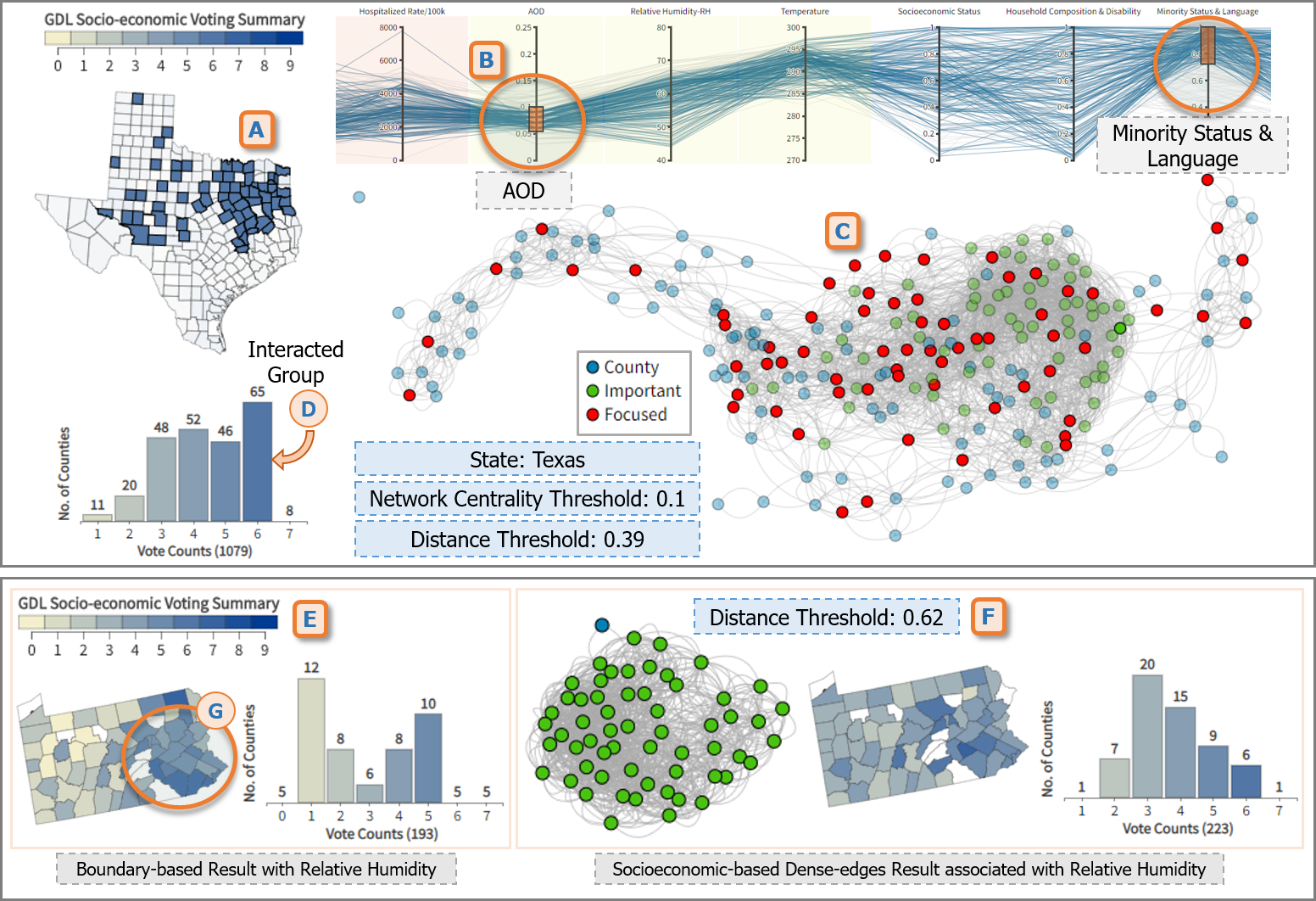}
    \caption{Coordinated multi-views demonstrating the interactive features in sense-making the GDL model result for TX and PA when associated with AOD and relative humidity.}
    \label{fig:use-case-1}
\end{figure}

\textbf{Improved forecasting in TX due to AOD: }
In Fig. \ref{fig:use-case-1}A, the choropleth map at the county level of TX illustrates North Eastern and Great Houston regions are better predicted, by multiple DL models at a $\alpha=0.1$ significant level, when introducing measurements of particles of the atmosphere. This fact can be associated with a higher amount of particles in the air of these counties when contrasted with their average rates. As an example, the on-road air pollution of Houston is 9 times greater than its metro-area counterparts~\cite{ANT}. Additionally, a view of the Overall Vulnerability Index~\cite{Web:ATSDR} shows a significant relationship between counties with the highest impact of air quality on hospitalizations and those socioeconomically vulnerable counties as shown in Fig. \ref{fig:use-case-1}B. Further uncovering the structural inequalities and disparities, in health access and health outcomes, of socioeconomically disadvantaged communities (Fig. \ref{fig:use-case-1}C).


\textbf{Linked racial disparities with clinical severity in PA:}
Forecasting of COVID-19 hospitalizations in Southeastern PA gets improved when associated with daily rates of relative humidity. This further supports the health impact of pollution on these communities when linked with the fact that PA ranks as one of the 25 worst US metro areas for ozone and year-round particle pollution~\cite{ALA}. There is a further significant improvement, in the total number of votes, when adding dense socioeconomic-based counties' connectivity (223) against the boundary-based network (193). Even more, a visual comparison between our consensus model and the Minority Status \& Language index shows a compelling relationship between racial disparities and COVID-19 clinical severity. Thus showing the important role that environmental (in)justice is playing to prevent COVID-19 clinical severity in PA, and across communities around the world~\cite{EnvJusRoundtable:Sacoby:2020}.

\begin{figure}[t]
    \centering
    \includegraphics[width=\linewidth]{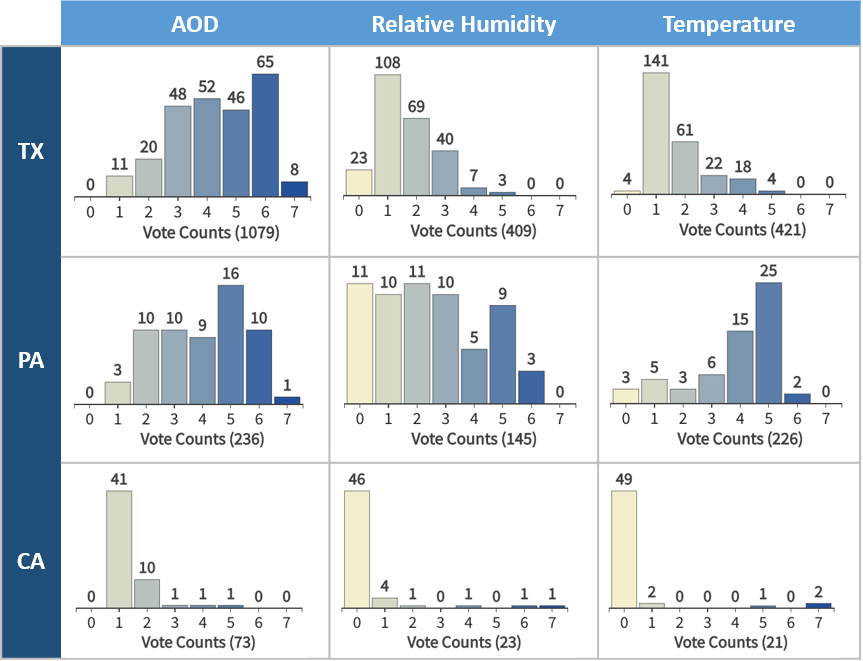}
    \caption{Comparing the GDL voting results where counties are grouped by vote counts for TX, PA, and CA to learn the impact of AOD, relative humidity, and temperature in COVID-19 clinical severity.}
    \label{fig:voting-comparison}
\end{figure}

\textbf{State level impact of atmospheric factors:}
%
The statistically verified votes in our consensus model allow us to identify regions where specific atmospheric factors could explain COVID-19 clinical severity. 
At a state level, we can conclude that AOD plays a bigger role than relative humidity and temperature when linked to forecasting hospitalization numbers, conditioned to multiple socioeconomic indices. Fig. \ref{fig:voting-comparison} presents the total number and distribution of votes in the three states when adding each atmospheric variable AOD (TX: 1079, PA: 236, CA: 73), relative humidity (TX: 409, PA: 145, CA: 23), and temperature (TX: 421, PA: 226, CA: 21). These findings further support the current trend of rethinking our engagement with the natural world, mainly because of our role as generators of particles in the air, but also from a wider environmental (in)justice perspective due to the impact in already vulnerable communities~\cite{EnvJusRoundtable:Sacoby:2020}.

\subsection{Discussion and Limitations}

The proposed visualization interface offers a number of multifold perspectives for health analytics way beyond the focus of COVID-19 disease tracking. First, the developed interface allows public health practitioners to perform retrospective spatiotemporal analysis of potential atmospheric contributors to clinical severity and mortality associated with respiratory diseases at various time periods and scales, while systematically accounting for socio-demographic information and quantifying uncertainties. As a result, such analysis can shed light on, for example, the potential (dis)similarities between the emerging and earlier recorded variants of a virus and the associated sensitivity of these variants to weather and climate. Second, coupled with weather forecasting, the visualized consensus among GDL models can be used to identify areas of the highest vulnerability risks and potential next hotspots requiring proactive medical emergency readiness. Third, more generally, armed with the new visualization interface, public policymakers can investigate such questions of critical societal importance as climate justice and their impacts on various disease progression stages. 

While the interactive visualization interface provides coordinated visualizations and rich user interactions that support primary and resulting data exploration and facilitate sense-making, it relies on pre-computed model results. Hence, the user is responsible to run their own machine-learning workflow to visually analyze data in the proposed interface. In our future work, we will address this concern by leveraging a visual analytics framework where the user is able to create workflows, prepare analytical scripts, and execute from the interface which would essentially end up utilizing the proposed interactive visualizations \cite{Nayeem2021visual}. A visual analytics system will also facilitate the user to interactively experiment with the parameters and statistical significance. Moreover, we plan to conduct a user study with domain experts to report a quantitative evaluation of our proposed visualization system.


\section{Conclusion}
The COVID-19 pandemic provides a unique opportunity to study the interlinks between environment, climate (in)justice, and biosurveillance. From a contemporary AI approach, this paper aims to add up to the collective knowledge of the lessons learned to identify and tackle structural inequalities contributing to higher mortality in vulnerable communities, and its relationships with atmospheric factors. We introduce an interactive interface and consensus analysis with coordinated multi-views that supports analyzing the impacts of atmospheric observations and socioeconomic factors on COVID-19 clinical severity. We present three usage scenarios from the proposed interactive interface to demonstrate the usability and its scientific value in analyzing COVID-19 clinical severity. Our findings support the impact of aerosols on the Earth's radiation budget and air quality, and its consequences for health fairness across communities of different socioeconomic backgrounds. Since our machine learning and interactive framework leverages NASA's satellite-based observations, our consensus analysis can be extended to a wider coverage over the entire globe. 

\section*{Acknowledgement}
This project is partially supported by the grants NASA 21-AIST21\_2-0059 and NSF Data Infrastructure Building Blocks (DIBBs) Program (Award \#1640818).
\scriptsize{
\bibliographystyle{IEEEtran}
\bibliography{main}
}

\end{document}